\documentclass[11pt,a4paper]{article}
\usepackage{jcappub}

\usepackage{url}
\usepackage{lscape}
\usepackage{threeparttable}

\def\beq{\begin{equation}}
\def\eeq{\end{equation}}
\def\alwaysmath#1{{\ifmmode{#1}\else{$#1$}\fi}}
\def\he#1{\hbox{\alwaysmath{{}^{#1}}{\rm He}}}
\def\li#1{\hbox{\alwaysmath{{}^{#1}}{\rm Li}}}

\def\10830{{He~I $\lambda$10830}}

\title{
The effects of \10830 on helium abundance determinations
}

\author[a]{Erik Aver}
\author[b,c,d]{Keith~A.~Olive}
\author[b,d]{Evan~D.~Skillman}

\affiliation[a]{Department of Physics, Gonzaga University, \\
502 E Boone Ave, Spokane, WA 99258}
\emailAdd{aver@gonzaga.edu}
\affiliation[b]{School of Physics and Astronomy, University of Minnesota, \\
116 Church St. SE, Minneapolis, MN 55455}
\affiliation[c]{William I. Fine Theoretical Physics Institute, University of Minnesota, \\
116 Church St. SE, Minneapolis, MN 55455}
\emailAdd{olive@umn.edu}
\affiliation[d]{Minnesota Institute for Astrophysics, University of Minnesota, \\
116 Church St. SE, Minneapolis, MN 55455}
\emailAdd{skillman@astro.umn.edu}

\abstract{
Observations of helium and hydrogen emission lines from metal-poor extragalactic H~II regions, combined with estimates of metallicity, provide an independent method for determining the primordial helium abundance, Y$_{p}$.  Traditionally, the emission lines employed are in the visible wavelength range, and the number of suitable lines is limited.  Furthermore, when using these lines, large systematic uncertainties in helium abundance determinations arise due to the degeneracy of physical parameters, such as temperature and density.  Recently, Izotov, Thuan, \& Guseva (2014) have 
pioneered adding the \10830 infrared emission line in helium abundance determinations.  The strong electron density dependence of \10830 makes it ideal for better constraining density, potentially breaking the degeneracy with temperature.  We revisit our analysis of the dataset published by Izotov, Thuan, \& Stasi\'nska (2007) and incorporate the newly available observations of \10830 by scaling them using the observed-to-theoretical Paschen-gamma ratio.  The solutions are better constrained, in particular for electron density, temperature, and the neutral hydrogen fraction, improving the model fit to data, with the result that more spectra now pass screening for quality and reliability, in addition to a standard 95\% confidence level cut.  Furthermore, the addition of \10830 decreases the uncertainty on the helium abundance for all galaxies, with reductions in the uncertainty ranging from 10-80\%.  Overall, we find a reduction in the uncertainty on Y$_{p}$ by over 50\%.  From a regression to zero metallicity, we determine Y$_{p}$ $=$ 0.2449 $\pm$ 0.0040, consistent with the BBN result, Y$_{p}$ $=$ 0.2470 $\pm$ 0.0002, based on the Planck determination of the baryon density.  The dramatic improvement in the uncertainty from incorporating \10830 strongly supports the case for simultaneous (thus not requiring scaling) observations of visible and infrared helium emission line spectra.  
}


\arxivnumber{}

\begin{document}

\begin{flushright}UMN-TH-3425/15\\FTPI-MINN-15/11\\
March 2015\end{flushright}

\maketitle
\flushbottom

\section{Introduction}
Although the determination of the fundamental parameters of the standard cosmological model including dark matter and dark energy, $\Lambda$CDM, by WMAP \citep{wmap, wmap2} and Planck \citep{planck, planck15} are unparalleled, big bang nucleosynthesis and the observations of the light element abundances offer an important cross-check, in particular on the determination of the baryon density.  The most recent Planck result \citep{planck15} for the baryon density, $\Omega_B h^2 = 0.02226 \pm 0.00016$, corresponds to a baryon-to-photon ratio of $\eta = (6.10 \pm 0.04) \times 10^{-10}$.  Because the uncertainty in $\eta$ is now less than 1\%, standard big bang nucleosynthesis (SBBN) \citep{wssok,osw,fs} is a parameter-free theory \citep{cfo2}, and relatively precise predictions of the primordial abundances of the light elements D, $^{3}$He, $^{4}$He, and $^{7}$Li are available \citep{cfo,coc,cfo3,coc2,cyburt,coc3,cuoco,serp,cfo5,pis,coc4,cuv,cuv2,cfoy}.  While the \li7 abundance remains problematic \citep{cfo5}, recent D/H determinations from quasar absorption systems have become quite precise, in their own right, and they are in excellent agreement with the prediction from SBBN and the CMB \citep{dh}. Using a neutron mean life of $880.3 \pm 1.1$ s \citep{rpp}, SBBN yields a primordial abundance for $^{4}$He, $Y_{p}$, of $Y_p = 0.2471 \pm 0.0002$, using the Planck determined value of $\eta$ \citep{cfoy}.  By allowing $Y_{p}$ to vary as an independent parameter, fits to CMB anisotropies allow for a determination of $Y_p$ within the context of $\Lambda$CDM.  The recent Planck results found $Y_p = 0.251 \pm 0.027$ \cite{planck15}.  Fortunately, the helium abundance from emission line measurements provide significantly better precision.

To test SBBN beyond D/H, it is clear that precise determinations of \he4 are necessary.  In addition, \he4 still provides important constraints on the physics of the early universe beyond the standard model \citep{cfos}.  Nevertheless, obtaining better than 1\% precision for individual objects remains a challenge.  \he4 abundance determinations are generally fraught with systematic uncertainties  and degeneracies among the input parameters needed to model emission line fluxes \citep{os01,os04,its07}. The primordial abundance of \he4 is determined by fitting the helium abundance versus metallicity (oxygen), and extrapolating back to very low metallicity \citep{ptp74}.  

Our method for determining the \he4 abundance in an individual H~II region is based on a Markov Chain Monte Carlo (MCMC) analysis \citep{AOS2,AOS3,AOPS}. Using calculated emissivities \citep{pfsd,pfsdc}, we model emission fluxes for a number of H and He lines relative to H$\beta$.  The model depends on several physical parameters associated with the H~II region: electron density, $n_e$, temperature, $T$, optical depth $\tau$, underlying stellar H and He absorption, $a_H$ and $a_{He}$, reddening, $C(H\beta$), the fraction of neutral hydrogen, $\xi$, and, of course, the He abundance, parametrized in terms of the abundance by number (relative to H) of ionized He, $y^+$.  MCMC scans of our 8-dimensional parameter space map out the likelihood function based on the $\chi^2$ given by
\beq
\chi^2 = \sum_{\lambda} {(\frac{F(\lambda)}{F(H\beta)} - {\frac{F(\lambda)}{F(H\beta)}}_{meas})^2 \over \sigma(\lambda)^2},
\label{eq:X2}
\eeq
where the emission line fluxes, $F(\lambda)$, are measured and calculated for six helium lines ($\lambda\lambda$3889, 4026, 4471, 5876, 6678, and 7065) and three hydrogen lines (H$\alpha$, H$\gamma$, H$\delta$), each relative to H$\beta$.  The $\chi^{2}$ in eq.\ \ref{eq:X2} runs over all He and H lines and $\sigma(\lambda)$ is the measured uncertainty in the flux ratio at each wavelength. 

We also adopt a weak prior on the temperature based on the temperature derived from [O~III] emission lines. $\chi^2_{T}=(T-T(OIII))^2/\sigma^2)$ is added to eq.\ \ref{eq:X2}, where we adopt $\sigma = 0.2\,T_{meas}$, which is much greater than the estimated temperature difference of 11\% between the measurements of temperature from ionized helium and doubly ionized oxygen lines \citep{ppl02,guseva06,guseva07}. For more detail on the applied temperature prior see \citet[][AOS2]{AOS2}. Here we also discuss the implications of removing the prior. 

Once minimized, best-fit solutions for the eight physical parameter inputs, including $y^+$, are found, and uncertainties in each quantity can be obtained by calculating a 1D marginalized likelihood. With eight parameters and only nine observables, our model is nearly under-determined.  
 
This MCMC analysis was applied to the large HeBCD dataset of \citet[][ITS07]{its07} in \citet[][AOS3]{AOS3}.  As discussed in AOS3, we consider only those objects with measured He~I $\lambda4026$ emission lines.  We then select those objects with $\chi^{2}<4$, corresponding to a standard $\sim$95\% confidence level for one degree of freedom.  Using a revised set of emissivities \citep{pfsd,pfsdc}, in our most recent work, \citet[][AOPS]{AOPS}, we found only 16 objects which satisfy this criterion and could be used to extrapolate to zero metallicity and determine $Y_{p}$. Due to the large uncertainties in $y^+$ found for each object, and the relatively small number of objects for which the model acceptably describes the data, the uncertainty on our determination of $Y_{p}$ was not much better than 4\%. 

Recently, \citet[][ITG14]{itg14} presented data for the near-infrared \10830 emission line for a large sample of metal-poor nebulae and pioneered the addition of this line in helium abundance analyses.  While not only providing an additional test for the model, this particular line shows a strong sensitivity to the electron density.  It can, therefore, be very useful in breaking the degeneracy between density and temperature \citep{os04, AOS}, allowing for more accurate determinations of $y^+$ in each region.  Here, we make use of these new observations in a new MCMC analysis.  We consider only observations for which all seven He lines have reported values in ITS07 \& ITG14 \citep{its07,itg14}.  While the initial data set is reduced (not every object measured in ITS07 has a measured \10830 emission line in ITG14 and vice versa), better constrained parameters allow more objects to yield statistically reliable and physically meaningful solutions.  
As a result, we base our current analysis on 15 objects (with some, but not full, overlap with the sample used in AOPS \citep{AOPS}). The uncertainty in $y^+$ in individual objects is, in fact, significantly reduced (by as much as 80\%) and the extrapolated primordial abundance is now determined to 1.6\%: $Y_p = 0.2449 \pm 0.0040$.

This paper is organized as follows.
First, \S \ref{10830} introduces \10830, with particular emphasis on its electron density dependence.  Second, in \S \ref{Synthetic}, Monte Carlo testing with synthetic data including \10830 is carried out to examine its diagnostic power.  This includes examining cases both with and without the inclusion of an O[III] temperature prior.  Third, \S \ref{ITG14} offers a brief overview of the ITG14 observations and dataset, along with the required scaling of the \10830 observations.  In \S \ref{Sample}, the set of ITS07 galaxies in ITG14 are analyzed, demonstrating the impact of \10830 on the solutions.  Subsequently, $Y_p$ is determined in \S \ref{Results}.  Finally, \S \ref{Conclusion} offers a discussion of the results, possible systematic effects or additional uncertainties, and the benefits of further observations with \10830.

\begin{figure}
\centering  
\resizebox{\textwidth}{!}{\includegraphics{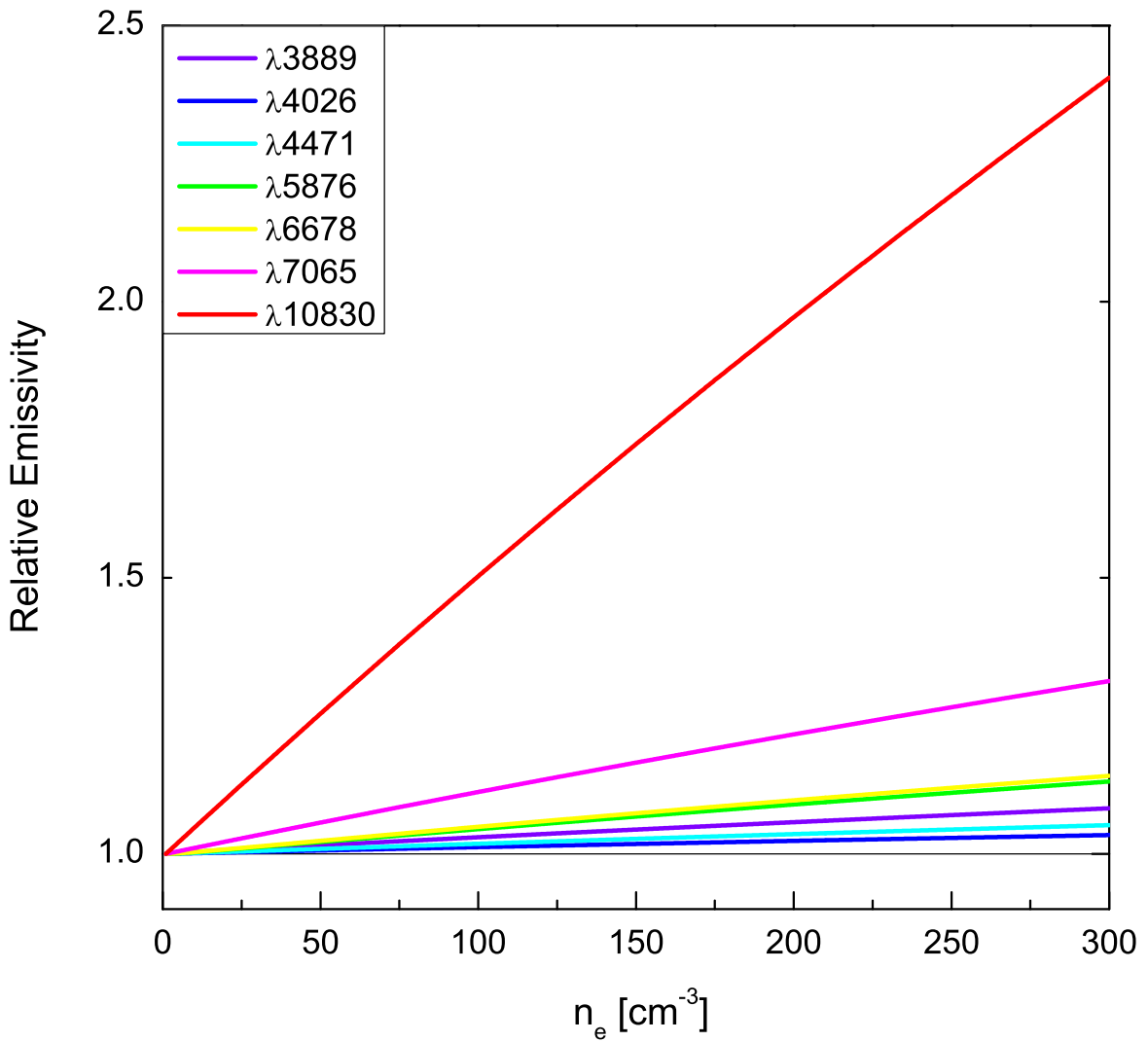}}
\caption{
Comparison of the \citet{pfsdc} emissivities, including the collisional correction, versus density, using a temperature of 18,000~K, for He~I $\lambda\lambda$3889, 4026, 4471, 5876, 6678, 7065, 10830.  The much stronger density dependence of \10830 is apparent.  
}
\label{Emissivity}
\end{figure}

\begin{figure}
\centering  
\resizebox{\textwidth}{!}{\includegraphics{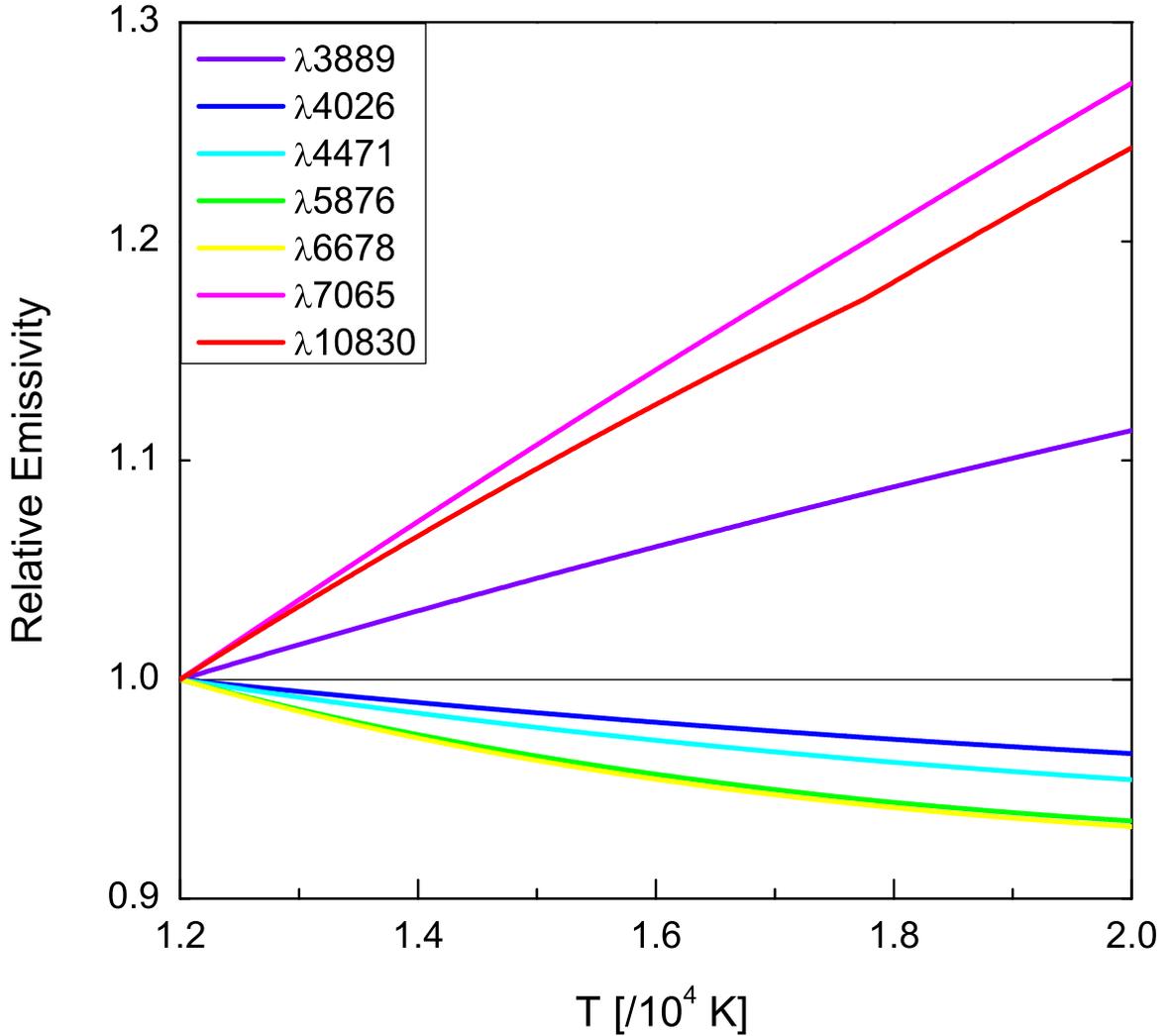}}
\caption{
Comparison of the \citet{pfsdc} emissivities, including the collisional correction, versus temperature, using a density of 100 cm$^{-3}$, for He~I $\lambda\lambda$3889, 4026, 4471, 5876, 6678, 7065, 10830.  
}
\label{Emissivity_T}
\end{figure}

\section{The diagnostic power of the infrared He emission line, $\lambda$10380} \label{10830}

\10830 is emitted in a $2\,^{3}\!P$ to $2\,^{3}\!S$ transition ($\Delta\ell=1$).  Corresponding to its small energy change ($\Delta E=1.14$ eV), it is readily collisionally excited from the metastable $2\,^{3}\!S$ triplet state.  It is this large population due to collisional excitation that makes it so sensitive to the H~II region's density.  Figure \ref{Emissivity} shows the sensitivity to density for all of the He emission lines considered. The figure clearly shows the dramatically stronger density dependence of \10830, with a slope over four times greater than that of the next most sensitive line, He~I $\lambda$7065.  Therefore, we expect that \10830 should be an excellent density diagnostic, greatly constraining the density in our solutions.  Figure \ref{Emissivity_T} shows the temperature dependence of the He emission lines considered.  \10830 does not stand out, as it did in figure \ref{Emissivity}, but its temperature dependence is the second strongest, and, furthermore, its increase with temperature helps to balance out the three strong optical lines, He~I $\lambda\lambda$4471, 5876, \& 6678, all of whose emissivities decrease similarly with temperature.  

The He line flux ratios (compared to H$\beta$) used in eq.\ (\ref{eq:X2}) are calculated using 
\beq
\frac{F(\lambda)}{F(H\beta)}= y^{+}\frac{{E}(\lambda)}{E(H\beta)}{\frac{W(H\beta)+a_{H}(H\beta)}{W(H\beta)} \over \frac{W(\lambda)+a_{He}(\lambda)}{W(\lambda)}}{f_{\tau}(\lambda)}\frac{1}{1+\frac{C}{R}(H\beta)}10^{-f(\lambda)C(H\beta)} \, ,
\label{eq:F_He}
\eeq
where $W(\lambda)$ is the measured equivalent width, and two parameters, $a_H$ and $a_{He}$, characterize the wavelength-dependent underlying absorption for H and He, respectively. The function $f_\tau(\lambda)$ represents a correction for radiative transfer and depends on on $\tau$, $n_e$, and $T$. The  emissivity, $E$, including the He collisional corrections, is a function of $n_e$, and $T$.  The hydrogen collisional corrections are accounted for by $\frac{C}{R}$, which depends on the fraction of neutral-to-ionized hydrogen, $\xi$.  The final term in  eq.\ (\ref{eq:F_He}) accounts for reddening, $C(H\beta)$.  See AOS \& AOS2 \citep[][]{AOS, AOS2} for full details on the model and analysis method.  

To incorporate \10830 into our analysis, the following data are used.  The emissivities, including the collisional correction, are adopted from the recent work of \citet{pfsdc}, as was implemented in AOPS \citep{AOPS}, and  include \10830.  The radiative transfer equations in terms of optical depth come from the work of \citet{bss02}.  That work does not include a fitting formula for \10830, however.  Instead, the 10-level numerical calculation program was graciously provided by Bob Benjamin (private communication).  Fitting the same functional form to the \10830 optical depth data over the same range (n$_e$ = 1-300 cm$^{-3}$ and T = 1.2-2.0 x 10$^4$ K), we obtain:  
\beq
f_{\tau}(10830) = 1 + (\tau/2) [0.0149+(4.45 \times 10^{-3}-6.34\times 10^{-5}n_{e}+9.20 \times 10^{-8}{n_{e}}^2)T)],
\label{eq:f_tau}
\eeq
with a maximum fit error of 1.3\% (similar to the maximum fit errors reported in \citep{bss02}).  Correcting for the effects of underlying absorption for \10830 is more tenuous since the available data is very limited.  The 3 B-star supergiant spectra (where helium absorption is more significant) reported in \citet{ch99} show equivalent width (EW) underlying absorption of $\sim$0.5-2.0 \AA.  Because the EW of \10830 is large (typically 100-400 \AA), the correction of underlying absorption will be reassuringly small.  For this work, we adopt an underlying absorption coefficient of 0.8, relative to He~I $\lambda$4471 (i.e., He~I $\lambda$4471 carries a coefficient of 1.0), such that the underlying absorption correction applied will be $0.8 \times a_{He}$, where $a_{He}$ is the solution for underlying absorption (see AOS for further details \citep{AOS})).  This ratio of 0.8 for underlying absorption of \10830 relative to that of He~I $\lambda$4471 is the same as in adopted by ITG14 \citep{itg14} (though the approaches to determining the underlying absorption differ).

\section{Characterizing the impact of He~I $\lambda$10380 with synthetic testing} \label{Synthetic}

\begin{figure}[ht!]
\centering  
\resizebox{0.90\textwidth}{!}{\includegraphics{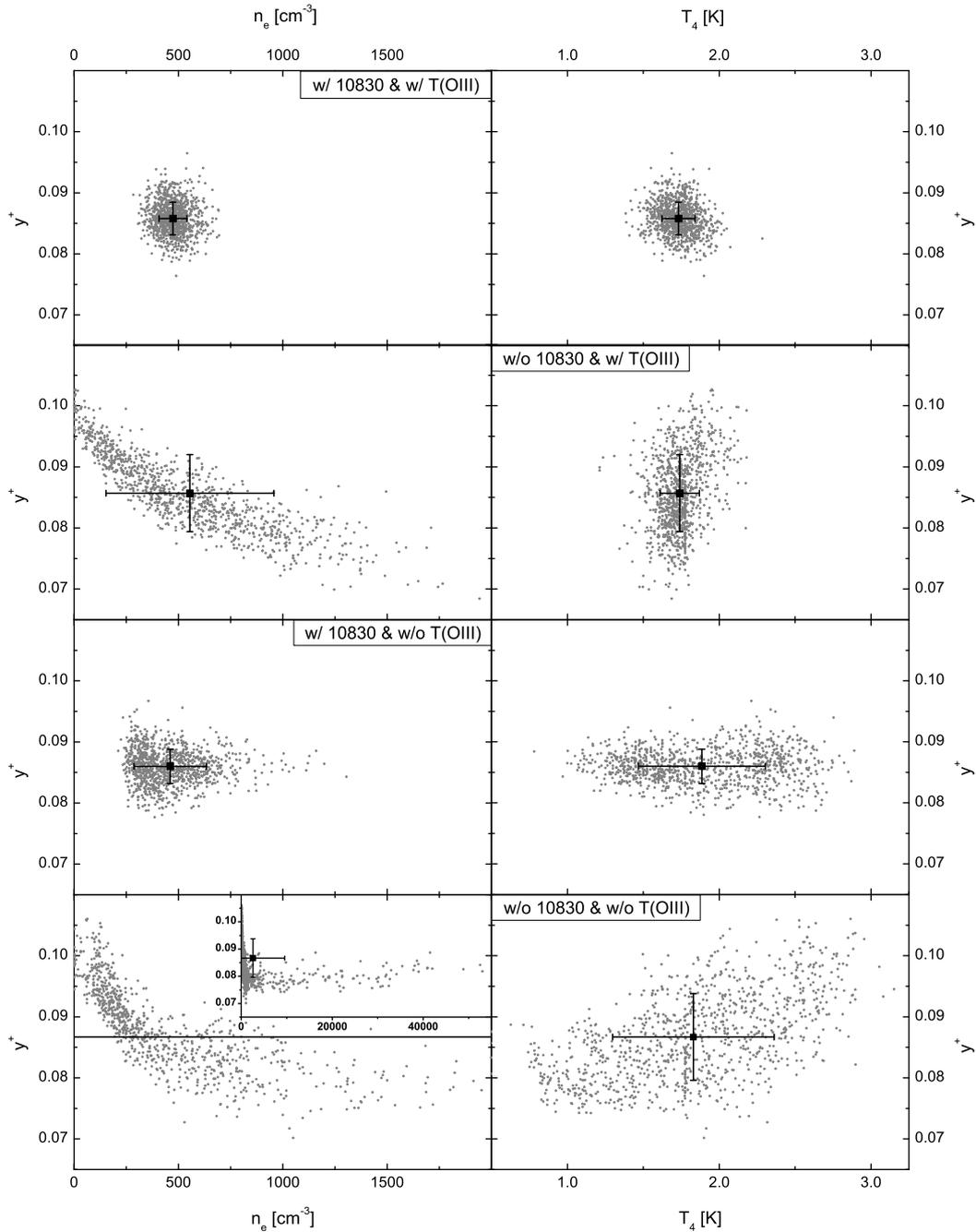}}
\caption{
Comparison of 1000 MC solutions for analysis with and without \10830, as well as with and without the [O~III] temperature prior.  For each of these four cases, the helium abundance is plotted separately versus density and temperature.  On each of these eight panels the average MC best-fit value is plotted with its dispersion.   From top to bottom the progression of the panels is as labeled:  w/ \10830 \& w/ T(O~III), w/o \10830 \& w/ T(O~III), w/ \10830 \& w/o T(O~III), and w/o \10830 \& w/o T(O~III).  The panels on the left are $y^+$ vs. $n_e$, while those on the right are $y^+$ vs.\ $T$, and, except for the inset in the bottom left panel, all of the axes are shared and show the same domain and range.  
}
\label{SynMC_DT}
\end{figure}

To begin discerning the effect and potential benefits of \10830, we conduct Monte Carlo analyses using synthetic data, since it offers a controlled environment for clear comparisons.  First, a set of input parameters were used to generate a synthetic spectrum.  These are given in the second column of table \ref{table:MC10000}. Then, 1000 sets of Gaussian perturbed fluxes were calculated with spreads of  3\% and 1.5\% of the unperturbed helium and hydrogen fluxes, respectively.  The best-fit solution (via $\chi^2$ minimization, see AOS \citep{AOS}) of each of these 1000 perturbed spectra was then found.  Finally, the average value and dispersion of those 1000 solutions was calculated for each parameter.  To allow comparisons showing the effects of \10830 and the [O~III] temperature prior, this process was repeated for the following four cases:  analysis with \10830 \& with T(O~III) included, without \10830 but with T(O~III), with \10830 but without T(O~III), and without \10830 \& without T(O~III).  The results of these MC simulations for $y^{+}$ vs.\ both $n_e$ and $T$ (the electron density and temperature solved for from the helium and hydrogen emission lines) are shown in figure \ref{SynMC_DT}.  The average value and dispersion for each of the four cases is provided in table \ref{table:MC10000}.  

\begin{table}[h!]
\centering
\vskip .1in
\begin{tabular}{lccccc}
\hline\hline
				&	 & w/ T(O~III) \&	  &			&			 & w/o T(O~III) \&  \\
				& Input	 & w/ \10830 			& w/ T(O~III) 	& w/ \10830 		& w/o \10830  \\
\hline
y$^+$				& 0.085	 & 0.0858 $\pm$ 0.0027	& 0.0857 $\pm$  0.0063	& 0.0860 $\pm$  0.0028	& 0.0867 $\pm$  0.0071 \\
n$_e$				& 500.0  & 473 $\pm$ 67		& 555 $\pm$ 402		& 460 $\pm$ 174		& 2570 $^{+6980}_{-2570}$ \\
a$_{He}$			& 0.5	 & 0.52 $\pm$  0.11	& 0.51 $\pm$  0.15	& 0.51 $\pm$  0.12	& 0.53 $\pm$  0.14 \\
$\tau$				& 1.0    & 0.78 $\pm$  0.31	& 0.80 $\pm$  0.73	& 0.62 $\pm$  0.56	& 0.52 $\pm$  0.72 \\
T$_e$				& 16,000 & 17,320 $\pm$ 1090	& 17,400 $\pm$ 1300	& 18,860 $\pm$ 4190	& 18,300 $\pm$ 5320 \\
C(H$\beta$)			& 0.1    & 0.09 $\pm$  0.03	& 0.09 $\pm$  0.03	& 0.09 $\pm$  0.03	& 0.08 $\pm$  0.04 \\
a$_{H}$				& 1.0    & 2.08 $\pm$  1.47	& 2.10 $\pm$  1.70	& 2.61 $\pm$  2.09	& 2.69 $\pm$  2.15 \\
$\xi$ $\times$ 10$^4$   	& 1.0    & 13 $^{+19}_{-13}$	& 14 $^{+31}_{-14}$	& 1040 $^{+31,620}_{-1040}$	& 8570 $^{+68530}_{-8570}$ \\
T(O~III)			& 17,000 &			&			&			& \\
\hline
\end{tabular}
\caption{Comparison of MC distributions with synthetic data}
\label{table:MC10000}
\end{table}

When the [O~III] prior and \10830 are included, the input data are nicely reproduced. As the figure and table clearly show, \10830 dramatically reduces the uncertainty on the density (by over 75\%), as well as improving the average, both with and without the inclusion of the T(O~III) prior.  As a result of the degeneracy between temperature and density (see OS04 \& AOS \citep{os04, AOS}), the uncertainty on the temperature is also decreased.  Corresponding to the better constrained density and temperature, the precision on the helium abundance, $y^+$, is greatly increased.  The inclusion of \10830 decreases the uncertainty on $y^+$ by over 50\%, when either the T(O~III) prior is or is not included.  In fact, in terms of the uncertainty on $y^+$, the inclusion of \10830 mostly obviates the benefit of the T(O~III) prior.  However, the T(O~III) prior still has a large effect in improving the temperature determination.  

The neutral hydrogen fraction, $\xi$, is very poorly constrained for very low temperatures, due to its exponential temperature dependence (see AOS \citep{AOS}).  As a result, the T(O~III) prior's exclusion of the lowest temperatures is crucial in restraining the range of MC solutions for $\xi$.  Objects with completely unphysical best-fit solutions for $\xi$ (i.e., $>$25\% neutral hydrogen) are excluded (see AOS3 \citep{AOS3}).  Therefore, the T(O~III) prior still offers potential utility in our analysis here, even if it does not improve the determination of $y^+$ when \10830 is included.  It should be noted that we first implemented T(O~III) as a weak prior to rule out unphysical double minima in some solutions (see AOS2 \citep{AOS2}).  The much better constrained density in the presence of \10830 has a similar effect, thus allowing for the possibility of removing the non-parametric prior.  

As can be seen particularly in the lowest of the density panels in figure \ref{SynMC_DT}, some MC realizations have solutions for which the density increases to very large values.  This is a consequence of the temperature-density degeneracy, and as discussed above, the corresponding very low temperatures allow the neutral hydrogen fraction to increase almost without bound (see AOS for further discussion \citep{AOS}).  To see the extent of the range in density found in this case, we provide an inset in that figure.  As a result of the extended range in $n_e$, the MC average values are biased upward for $n_e$ and $\xi$, with a correspondingly large uncertainty, as can be seen in the inset and the last column of table \ref{table:MC10000}.  This is one of the drawbacks of Monte Carlo via flux perturbation, as was discussed in AOS \citep{AOS}.  The MCMC method introduced in AOS2 \citep{AOS2} is not sensitive to this same behavior.  Nevertheless, the distribution of perturbed solutions is very effective at illustrating the effects of the T(O~III) prior and of \10830.  

For a further demonstration of the effects of \10830, one of the flux perturbed synthetic spectra was analyzed using the MCMC analysis we employ to determine the solutions to the physical parameter set \citep{AOS2, AOS3, AOPS}.  Rather than examining the distribution of 1000 flux simulations, the spectrum, synthetic or otherwise, is treated as an observation, for which a Monte Carlo over the eight model parameters (y$^{+}$, n$_{e}$, a$_{He}$, $\tau$, T, C(H$\beta$), a$_{H}$, $\xi$) is performed to determine the uncertainties about the best-fit values.  For one of the flux perturbed synthetic spectra, the solution (with uncertainties) was found for the same four cases discussed above and is reported in table \ref{table:SynMCMC}.  The conclusions from the above analysis are reinforced, with \10830 tremendously improving the density determination, both the best-fit value and the uncertainty.  This, in turn, improves the helium abundance determination.  Comparing the solutions with and without \10830, the 68\% confidence level uncertainty range for the density is decreased by $\sim$75\% \& $\sim$80\%, for the cases with and without the T(O~III) prior, respectively, while the uncertainty in $y^+$ sees corresponding decreases of $\sim$40\% \& $\sim$50\%.  Again, the efficacy of \10830 in improving our solutions is impressive.  

\begin{table}[ht!]
\centering
\vskip .1in
\begin{tabular}{lccccc}
\hline\hline
				& 	 & w/ T(O~III) \&		& 	 			&  				& w/o T(O~III) \&        \\
				& Input	 & w/ \10830 			& w/ T(O~III) 			& w/ \10830 			& w/o \10830       \\
\hline
y$^+$				& 0.085	 & 0.0870 $^{+0.0036}_{-0.0017}$ & 0.0805 $^{+0.0056}_{-0.0031}$ & 0.0872 $^{+0.0037}_{-0.0015}$ & 0.0815 $^{+0.0062}_{-0.0040}$  \\
n$_e$				& 500.0  & 457 $^{+135}_{-117}$		& 995 $^{+681}_{-469}$ 		& 519 $^{+226}_{-250}$ 		& 747 $^{+2540}_{-355}$  \\
a$_{He}$			& 0.5	 & 0.48 $^{+0.12}_{-0.08}$	& 0.38 $^{+0.13}_{-0.10}$ 	& 0.50 $^{+0.10}_{-0.09}$ 	& 0.38 $^{+0.14}_{-0.10}$  \\
$\tau$				& 1.0    & 1.42 $^{+0.38}_{-0.68}$	& 0.00 $^{+0.61}_{-0.00}$ 	& 1.60 $^{+0.38}_{-1.02}$ 	& 0.00 $^{+0.62}_{-0.00}$  \\
T$_e$				& 16,000 & 15,400 $^{+4222}_{-2052}$	& 17,620 $^{+2310}_{-2870}$ 	& 14,210 $^{+6880}_{-2310}$ 	& 19,380 $^{+2520}_{-8980}$  \\
C(H$\beta$)			& 0.1    & 0.13 $^{+0.02}_{-0.04}$	& 0.08 $^{+0.03}_{-0.03}$ 	& 0.12 $^{+0.02}_{-0.03}$ 	& 0.08 $^{+0.03}_{-0.04}$  \\
a$_{H}$				& 1.0    & 0.10 $^{+3.07}_{-0.10}$	& 3.89 $^{+1.68}_{-2.11}$ 	& 0.00 $^{+3.13}_{-0.00}$ 	& 4.20 $^{+1.52}_{-3.91}$  \\
$\xi$ $\times$ 10$^4$   	& 1.0    & 0 $^{+112}_{-0}$		& 26 $^{+136}_{-26}$ 		& 0 $^{+370}_{-0}$ 		& 12 $^{+1024}_{-12}$  \\
$\chi^2$			&	 & 3.11				& 1.47				& 2.71				& 1.37 \\
T(O~III)			& 17,000 &				&				& 				&   \\
\hline
\end{tabular}
\caption{Comparison of MCMC solutions with synthetic data}
\label{table:SynMCMC}
\end{table}

\section{The ITG14 observations and dataset} \label{ITG14}

\citet[][ITG14]{itg14} have recently published their observations of helium abundances using \10830.  The ITG14 dataset consists of near-infrared measurements of \10830 and hydrogen emission line Paschen $\gamma$ $\lambda$10940 \AA~(P$\gamma$) from 45 H~II regions in 43 low metallicity galaxies. The majority of the observations, 33 of 45, were  made with the 3.5 m Apache Point Observatory and TripleSpec spectrograph; 8 observations were taken with the 8.4 m Large Binocular Telescope and Lucifer spectrograph, as it allows observations of fainter, often lower metallicity, galaxies; and finally, four observations were retrieved from the European South Observatory data archives (2 VLT \& 2 NTT).  ITG14 combine these near-infrared spectra with 75 optical spectra taken from those 43 galaxies, where some galaxies have multiple independent observations.
Because the near-infrared and optical spectra were taken with different telescopes, apertures, and at different times, the measured \10830 flux must be scaled to match the optical flux measurements.  Similar to ITG14 \citep{itg14}, we scale \10830 using the observed P$\gamma$ flux and the theoretical flux ratio for P$\gamma$ relative to H$\beta$.  

\begin{eqnarray}
\frac{F(\lambda10830)}{F(H\beta)} & = & \frac{F(\lambda10830)}{F(P\gamma)}\frac{F(P\gamma)}{F(H\beta)} \nonumber \\
\frac{F(\lambda10830)}{F(H\beta)} & = & \frac{F(\lambda10830)}{F(P\gamma)}\frac{{E}(P\gamma)}{E(H\beta)}{\frac{W(H\beta)+a_{H}(H\beta)}{W(H\beta)} \over \frac{W(P\gamma)+a_{H}(P\gamma)}{W(P\gamma)}}\frac{1+\frac{C}{R}(P\gamma)}{1+\frac{C}{R}(H\beta)}10^{-f(P\gamma)C(H\beta)}.
\label{eq:F_10830}
\end{eqnarray}

The theoretical flux ratio for P$\gamma$ relative to H$\beta$ is calculated analogously to eq.\ (\ref{eq:F_He}).  As for all of the hydrogen emission lines, the emissivity ratio for P$\gamma$ relative to H$\beta$ is taken from \citet{hs87}.  The collisional correction for P$\gamma$ is calculated, as in AOS \citep{AOS}, using collisional excitation rates from \citet{and02} (similar to H$\delta$) and recombination rates from \citet{hs87}.  The estimate of the underlying absorption correction for P$\gamma$ is potentially much more significant than for \10830 because it is an intrinsically weaker emission line and is constrained by the same lack of data as in the case of  \10830.  The same 3 B-star supergiants discussed in \S \ref{10830} show a similar $\sim$0.5-2.0 \AA\ EW of underlying absorption for P$\gamma$ \citep{ch99}.  For this work, we adopt an underlying absorption coefficient of 0.4, relative to H$\beta$ (i.e., H$\beta$ carries a coefficient of 1.0), such that the underlying absorption correction applied will be $0.4 \times a_{H}$, where $a_{H}$ is the solution for underlying absorption based on all of the hydrogen lines (see AOS for further details \citep{AOS}). 

Of the 45 H~II regions with measured \10830 emission lines reported in ITG14 \citep{itg14}, 26 are also present in the 93 observations comprising the HeBCD dataset of ITS07 \citep{its07}, which we have previously analyzed in AOS3 \& AOPS \citep{AOS3, AOPS}.  Of the 93 HeBCD objects, He~I $\lambda$4026 is detected for 70.  To reduce systematic uncertainty that may be introduced by the absence of He~I $\lambda$4026 (see AOS3 for further details \citep{AOS3}), objects where He~I $\lambda$4026 is not reported are excluded in our analysis.  Of the 26 H~II regions in the HeBCD for which \10830 is reported in ITG14, He~I $\lambda$4026 is detected for 22.  However, of those 22, several have multiple independent observations, increasing the total number of spectra with both \10830 and He~I $\lambda$4026 to 31.  Those 31 sets of observations, with optical spectra taken from ITS07 and near-infrared from ITG14, comprise our initial dataset for this work.  

\section{Tracking the effects of \10830} \label{Sample}

For the 31 HeBCD observations from ITS07 \citep{its07} with \10830 observations reported in ITG14 \citep{itg14}, the effects of including \10830 into our helium abundance analysis is examined for two cases.  First, we examine the effects of the addition of \10830 to the analysis of AOPS \citep{AOPS}.  As discussed earlier, in AOS2, AOS3, \& AOPS \citep{AOS2, AOS3, AOPS}, a temperature prior based on [O~III] emission lines was utilized to eliminate unphysical double minima and improve the parameter determination.  However, it is a non-parametric (i.e., not fully ``self-consistent'') portion of our model, and it introduces a small bias in the solution.  Due the inclusion of \10830, the T(O~III) prior is no longer needed to eliminate double minima, and synthetic testing in \S \ref{Synthetic} showed that its benefit in reducing parameter uncertainties was greatly reduced by the inclusion of \10830.  Therefore, we also examine the effects of removing the T(O~III) prior once \10830 is included.  

For each of these analyses, cuts on the dataset are made following AOS3 \citep{AOS3} (also employed in AOPS \citep{AOPS}).  Of primary interest is the standard 95\% confidence level cut.  In AOS3, there was one degree of freedom (9 fluxes relative to H$\beta$ and 8 model parameters).  However, with the addition of \10830, there are now two degrees of freedom in our analysis.  Therefore, a 95\% confidence level corresponds to $\chi^2<6$, and best-fit solutions with $\chi^2>6$ were excluded.  Furthermore, objects with unphysical physical parameters, specifically $\xi > 0.333$ ($>$25\% neutral hydrogen), were also excluded.  Of the remaining, qualifying objects, objects with large corrections for systematic effects were flagged to limit potential systematic errors introduced by uncertainties in the models.  Objects with $\tau > 4$ and $\xi > 0.01$, where the 1-$\sigma$ lower bound does not encompass $\xi=0.001$, were flagged.  

In the analysis of AOPS \citep{AOPS}, the 31 objects in our present dataset broke down as follows:  11 qualifying, 2 flagged (for $\tau>4$), and 18 excluded, 12 for $\chi^2>4$, 3 for $\xi > 0.333$, and 3 for both.  When \10830 is added, the 11 previously qualifying objects are retained as qualifying.  One of the flagged objects is now qualifying, the other is still flagged for $\tau$.  Six previously excluded objects are no longer excluded, four qualifying and two flagged (for $\xi$).  Thus, the number of qualifying objects increases from 11 to 16, while the total number of flagged objects increases by one (from 2 to 3).  If the T(O~III) prior is then removed, only three originally excluded objects are reclaimed, one qualifying, and two flagged (for $\xi$), and one originally flagged object is excluded (the other remains flagged for $\tau$).  As a result, the qualifying and flagged datasets each increase by only one, from 11 to 12 and 2 to 3, respectively.  As a final note on qualifying and flagged objects, two of the qualifying and one of the flagged observations in the analysis retaining the T(O~III) prior, SBS~0335-052E1, SBS~0335-052E3, \& SBS~0335-052E3, respectively, are independent observations of the same H~II region.  As such, they are combined by weighted average into a single object, reducing the number of qualifying objects from 16 to 15, and the distinct flagged objects from 3 to 2.  

\begin{figure}
\centering  
\resizebox{0.97\textwidth}{!}{\includegraphics{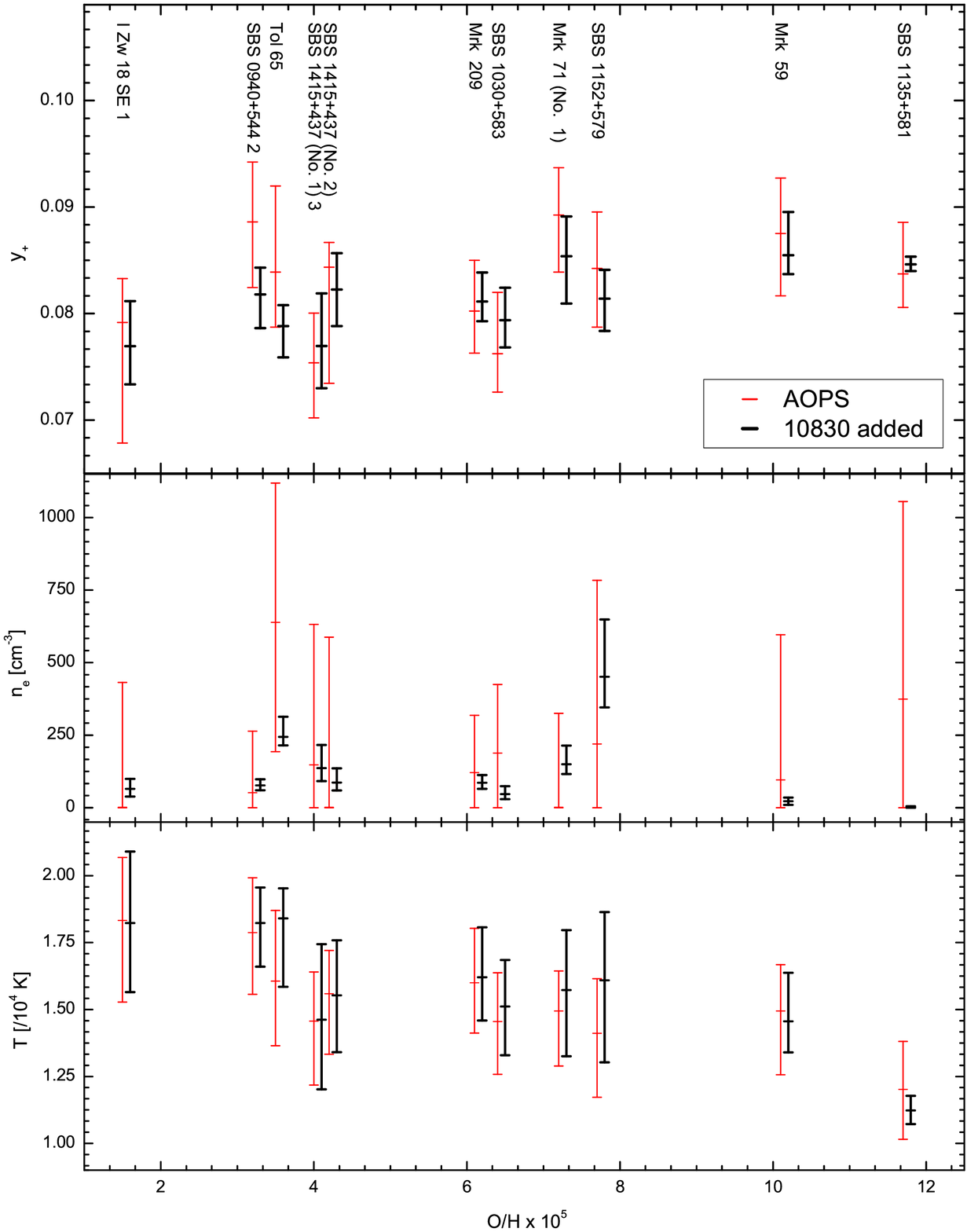}}
\caption{
Plot comparing the parameter solutions for $y^+$, $n_e$, and $T_4$ ($T_e/10^4$) for the 11 qualifying objects in AOPS with \10830 values in ITG14.  The lighter, thinner lines show the results given in AOPS.  The results after the inclusion of \10830 are given in the darker, thicker bars.  The uncertainty in $n_e$ is dramatically reduced by the addition of \10830 with a corresponding reduction in the uncertainty in $y^+$.
}
\label{AOPS-10830}
\end{figure}

\begin{figure}[ht!]
\centering  
\resizebox{0.97\textwidth}{!}{\includegraphics{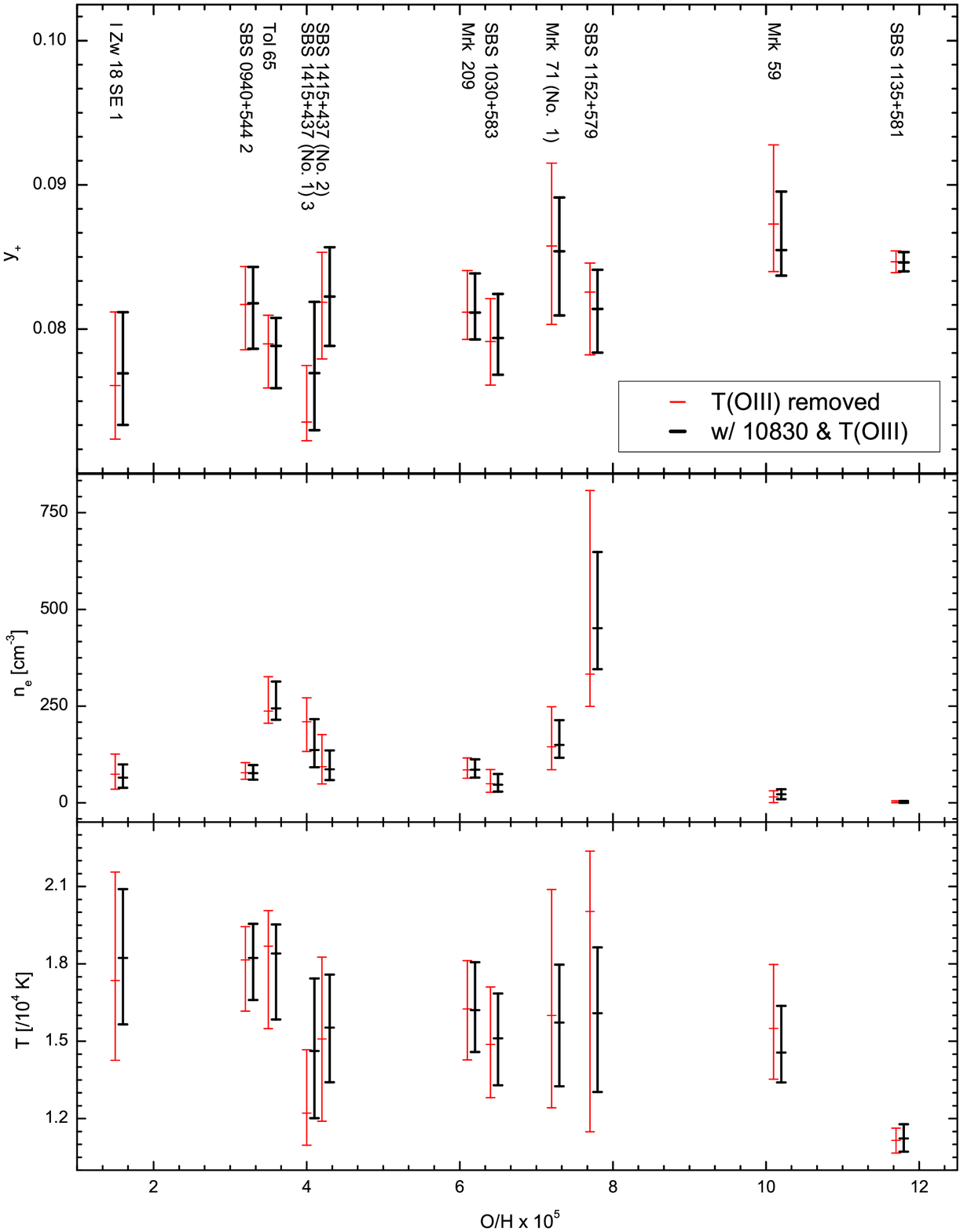}}
\caption{
Plot comparing the parameter solutions for $y^+$, $n_e$, and $T_4$ ($T_e/10^4$) for the 11 qualifying objects in AOPS with \10830 values in ITG14.  The lighter, thinner lines show the results where \10830 is included but the T(O~III) prior is removed.  The results including both \10830  and the T(O~III) prior are given in the darker, thicker bars.  Because including the T(O~III) prior does not significantly bias the values of $y^+$, and because its inclusion results in a greater yield of qualifying points, the T(O~III) prior is applied in the final analysis.  
}
\label{woTOIII-10830}
\end{figure}

For the 11 jointly qualifying objects in AOPS and this work, figures \ref{AOPS-10830} \& \ref{woTOIII-10830} compare the best-fit solutions and uncertainties for $y^+$, $n_e$, and $T_4$ ($T_e/10^4$), first for the addition of \10830, and then for the removal of the T(O~III) prior.  In figure \ref{AOPS-10830}, the effect of including \10830 is examined.  The AOPS values (whose analysis utilized the T(O~III) prior) are plotted adjacent to the results where \10830 is included (and the T(O~III) prior is retained).  The uncertainties for $y^+$ show decreases ranging from 10-80\% (see upper panel).  As was seen in \S \ref{Synthetic}, this substantial improvement in the determination of $y^+$ stems from the dramatically better constrained density.  The 68\% confidence level density range is reduced by over 60\%, with most objects seeing their uncertainty range on the density decrease by around 85\% (see middle panel).  Due to the temperature-density degeneracy (see OS04 \& AOS \citep{os04, AOS}), the better constrained density results in a reduced temperature uncertainty (see lower panel).  As density and temperature are the most important parameters in relating the helium abundance to the observed flux, $y^+$ is correspondingly better determined.  Note that in all cases, the new solutions are fully consistent with the prior solutions, though now with greatly reduced uncertainties. The best-fit values for $y^+$ decrease overall (4 of 11 increase, 7 decrease), but the average reduction in the helium abundance is small, approximately 2\%.  

Figure \ref{woTOIII-10830} shows the same parameter comparisons but examines the effect of removing T(O~III) prior after \10830 has been added.  Therefore, both series include \10830, and the results without and with the T(O~III) are plotted adjacent to each other.  Overall, there is a slight improvement in the parameter determinations from the inclusion of the T(O~III) prior, but it is not nearly as dramatic as in figure \ref{AOPS-10830}.  The uncertainties on $y^+$ generally decrease by 5-10\% due to the inclusion of the T(O~III) prior (see figure \ref{woTOIII-10830}, upper panel).  

Because the addition of \10830 increases the constraint on the density so strongly, the T(O~III) prior is no longer needed to eliminate unphysical double minima, while the T(O~III) prior's benefit in determining $y^+$ is greatly reduced (as was shown in \S \ref{Synthetic} and figure \ref{woTOIII-10830}).  However, as discussed in \S \ref{Synthetic}, the neutral hydrogen fraction, $\xi$, is very poorly constrained for low temperatures, due to its exponential temperature dependence (see AOS \citep{AOS}). As a result, the T(O~III) prior combined with \10830 helps to constrain the temperatures further than either does individually.  The preclusion of the lowest temperatures restricts $\xi$ and prevents the solution being excluded due to a completely unphysical best-fit value for $\xi$ ($>$25\% neutral hydrogen).  Therefore, because the T(O~III) prior allows three more objects to qualify, increasing the qualifying dataset by 25\% compared to adding \10830 but removing the T(O~III) prior, we favor retaining the T(O~III) prior for this analysis.  

This dataset of 15 qualifying objects and 2 flagged objects (analyzed including both \10830 \& the T(O~III) prior) comprises our Final Dataset.  These are the objects for which the model is a good fit and which returns physically meaningful parameter solutions.  As discussed above and as shown in figure \ref{AOPS-10830}, the parameter determinations of their solutions are significantly improved compared to our analysis in AOPS \citep{AOPS}, which did not include \10830.  The best-fit solutions and uncertainties of the Final Dataset are presented in table \ref{table:GTO}, and they are used to determine $Y_p$ in the following section (\S \ref{Results}).

\begin{landscape}
\begin{table}[b!]
\footnotesize
\centering
\vskip .1in
\begin{tabular}{lccccccccc}
\hline
\hline
Object 			& He$^+$/H$^+$     		   &  n$_e$     		      & a$_{He}$     		   & $\tau$     			& T$_e$     		   & C(H$\beta$)     	  	& a$_H$		     & $\xi$ $\times$ 10$^4$ 			& $\chi^2$ \\
\hline
\multicolumn{10}{c}{Final Dataset Not Flagged (Qualifying)} \\
\hline
I~Zw~18~SE~1    & 0.07693 $^{+0.00423}_{-0.00358}$ &      65 $^{+     34}_{-     26}$ &  0.19 $^{+ 0.21}_{- 0.19}$ &  0.31 $^{+ 0.71}_{- 0.31}$ & 18,227. $^{+2670.}_{-2575.}$ &  0.01 $^{+ 0.02}_{- 0.01}$ &  3.96 $^{+ 0.84}_{- 0.74}$ &        1 $^{+      13}_{-       1}$ &  0.4 \\
SBS~0335-052E1  & 0.07859 $^{+0.00418}_{-0.00470}$ &     154 $^{+     72}_{-     38}$ &  0.08 $^{+ 0.14}_{- 0.08}$ &  3.97 $^{+ 0.92}_{- 0.68}$ & 20,669. $^{+2213.}_{-3920.}$ &  0.09 $^{+ 0.02}_{- 0.03}$ &  3.33 $^{+ 1.44}_{- 1.65}$ &        6 $^{+      35}_{-       6}$ &  0.8 \\
SBS~0335-052E3  & 0.08443 $^{+0.00396}_{-0.00385}$ &     138 $^{+     53}_{-     28}$ &  0.32 $^{+ 0.08}_{- 0.08}$ &  2.39 $^{+ 0.74}_{- 0.67}$ & 21,780. $^{+3316.}_{-3807.}$ &  0.21 $^{+ 0.02}_{- 0.02}$ &  0.98 $^{+ 1.10}_{- 0.98}$ &        7 $^{+      19}_{-       5}$ &  4.4 \\
J0519+0007      & 0.08875 $^{+0.00461}_{-0.00402}$ &     675 $^{+    143}_{-    110}$ &  0.34 $^{+ 0.27}_{- 0.26}$ &  3.35 $^{+ 0.64}_{- 0.59}$ & 16,036. $^{+2738.}_{-2732.}$ &  0.16 $^{+ 0.04}_{- 0.04}$ &  0.00 $^{+ 0.72}_{- 0.00}$ &      235 $^{+    1532}_{-     235}$ &  4.3 \\
SBS~0940+544~2  & 0.08179 $^{+0.00252}_{-0.00316}$ &      77 $^{+     21}_{-     17}$ &  0.42 $^{+ 0.14}_{- 0.16}$ &  0.00 $^{+ 0.28}_{- 0.00}$ & 18,232. $^{+1324.}_{-1629.}$ &  0.06 $^{+ 0.02}_{- 0.02}$ &  2.62 $^{+ 1.02}_{- 1.14}$ &        6 $^{+      11}_{-       6}$ &  1.6 \\
Tol~65          & 0.07883 $^{+0.00195}_{-0.00294}$ &     244 $^{+     69}_{-     29}$ &  0.66 $^{+ 0.08}_{- 0.09}$ &  0.00 $^{+ 0.49}_{- 0.00}$ & 18,403. $^{+1124.}_{-2556.}$ &  0.10 $^{+ 0.02}_{- 0.02}$ &  4.70 $^{+ 0.76}_{- 1.03}$ &        6 $^{+      14}_{-       6}$ &  4.1 \\
SBS~1415+437~(No.~1)~3 & 0.07694 $^{+0.00494}_{-0.00396}$ &     136 $^{+     80}_{-     45}$ &  0.27 $^{+ 0.12}_{- 0.13}$ &  1.10 $^{+ 0.76}_{- 0.74}$ & 14,621. $^{+2817.}_{-2602.}$ &  0.11 $^{+ 0.03}_{- 0.03}$ &  0.82 $^{+ 1.18}_{- 0.82}$ &       56 $^{+     269}_{-      56}$ &  2.7 \\
SBS~1415+437~(No.~2) & 0.08226 $^{+0.00341}_{-0.00343}$ &      87 $^{+     49}_{-     28}$ &  0.48 $^{+ 0.09}_{- 0.09}$ &  0.98 $^{+ 0.65}_{- 0.63}$ & 15,531. $^{+2056.}_{-2122.}$ &  0.00 $^{+ 0.01}_{- 0.00}$ &  3.54 $^{+ 0.63}_{- 0.71}$ &        0 $^{+      41}_{-       0}$ &  1.4 \\
CGCG~007-025~(No.~2) & 0.08867 $^{+0.00462}_{-0.00623}$ &     181 $^{+    103}_{-     36}$ &  0.33 $^{+ 0.20}_{- 0.20}$ &  0.46 $^{+ 1.02}_{- 0.46}$ & 16,444. $^{+2449.}_{-3685.}$ &  0.11 $^{+ 0.04}_{- 0.04}$ &  3.01 $^{+ 1.64}_{- 2.20}$ &      126 $^{+    1303}_{-     126}$ &  1.8 \\
Mrk~209         & 0.08114 $^{+0.00272}_{-0.00186}$ &      86 $^{+     27}_{-     21}$ &  0.29 $^{+ 0.08}_{- 0.08}$ &  0.24 $^{+ 0.40}_{- 0.24}$ & 16,207. $^{+1862.}_{-1619.}$ &  0.00 $^{+ 0.02}_{- 0.00}$ &  2.59 $^{+ 0.85}_{- 0.82}$ &        0 $^{+      19}_{-       0}$ &  0.3 \\
SBS~1030+583    & 0.07937 $^{+0.00306}_{-0.00255}$ &      47 $^{+     28}_{-     18}$ &  0.22 $^{+ 0.07}_{- 0.08}$ &  0.31 $^{+ 0.50}_{- 0.31}$ & 15,114. $^{+1736.}_{-1820.}$ &  0.00 $^{+ 0.02}_{- 0.00}$ &  1.40 $^{+ 0.35}_{- 0.39}$ &        0 $^{+      49}_{-       0}$ &  1.6 \\
Mrk~71~(No.~1)  & 0.08539 $^{+0.00374}_{-0.00445}$ &     150 $^{+     64}_{-     33}$ &  0.49 $^{+ 0.11}_{- 0.13}$ &  1.25 $^{+ 0.71}_{- 0.55}$ & 15,724. $^{+2245.}_{-2469.}$ &  0.08 $^{+ 0.02}_{- 0.02}$ &  3.87 $^{+ 2.22}_{- 2.38}$ &       59 $^{+     214}_{-      59}$ &  2.5 \\
SBS~1152+579    & 0.08139 $^{+0.00272}_{-0.00303}$ &     452 $^{+    197}_{-    106}$ &  0.30 $^{+ 0.08}_{- 0.07}$ &  0.93 $^{+ 0.59}_{- 0.50}$ & 16,090. $^{+2547.}_{-3060.}$ &  0.17 $^{+ 0.03}_{- 0.02}$ &  4.26 $^{+ 1.16}_{- 1.34}$ &      100 $^{+     622}_{-     100}$ &  1.5 \\
Mrk~59          & 0.08548 $^{+0.00405}_{-0.00177}$ &      22 $^{+     13}_{-     13}$ &  0.52 $^{+ 0.07}_{- 0.05}$ &  0.73 $^{+ 0.34}_{- 0.44}$ & 14,558. $^{+1817.}_{-1158.}$ &  0.12 $^{+ 0.02}_{- 0.02}$ &  1.72 $^{+ 0.77}_{- 0.44}$ &        0 $^{+      62}_{-       0}$ &  0.7 \\
SBS~1135+581    & 0.08462 $^{+0.00072}_{-0.00063}$ &       1 $^{+      3}_{-      1}$ &  0.39 $^{+ 0.04}_{- 0.04}$ &  1.18 $^{+ 0.29}_{- 0.32}$ & 11,226. $^{+ 553.}_{- 508.}$ &  0.11 $^{+ 0.02}_{- 0.02}$ &  2.89 $^{+ 0.28}_{- 0.24}$ &        0 $^{+     472}_{-       0}$ &  4.9 \\
Mrk~450~(No.~1) & 0.08634 $^{+0.00441}_{-0.00335}$ &      97 $^{+     42}_{-     28}$ &  0.37 $^{+ 0.20}_{- 0.20}$ &  2.27 $^{+ 0.53}_{- 0.48}$ & 12,979. $^{+1321.}_{-1476.}$ &  0.15 $^{+ 0.03}_{- 0.03}$ &  2.31 $^{+ 1.62}_{- 1.65}$ &      171 $^{+     601}_{-     171}$ &  5.5 \\
\hline
\multicolumn{10}{c}{Final Dataset with Flags} \\
\hline
SBS~0335-052E2  & 0.08007 $^{+0.00407}_{-0.00301}$ &     136 $^{+     52}_{-     35}$ &  0.47 $^{+ 0.08}_{- 0.08}$ &  4.28 $^{+ 0.68}_{- 0.72}$ & 19,971. $^{+3345.}_{-2984.}$ &  0.04 $^{+ 0.02}_{- 0.02}$ &  3.11 $^{+ 1.18}_{- 0.99}$ &        2 $^{+       5}_{-       2}$ &  3.9 \\
Mrk~1315        & 0.09341 $^{+0.00274}_{-0.00184}$ &      10 $^{+     11}_{-     10}$ &  0.32 $^{+ 0.15}_{- 0.15}$ &  0.81 $^{+ 0.38}_{- 0.43}$ & 11,653. $^{+1218.}_{-1024.}$ &  0.11 $^{+ 0.02}_{- 0.02}$ &  0.00 $^{+ 1.04}_{- 0.00}$ &     3058 $^{+    7101}_{-    2276}$ &  4.5 \\
Mrk~1329        & 0.09060 $^{+0.00425}_{-0.00204}$ &     101 $^{+     34}_{-     34}$ &  0.25 $^{+ 0.15}_{- 0.11}$ &  1.03 $^{+ 0.42}_{- 0.44}$ & 11,223. $^{+1457.}_{-1006.}$ &  0.17 $^{+ 0.02}_{- 0.03}$ &  0.11 $^{+ 1.56}_{- 0.11}$ &     1613 $^{+    5435}_{-    1473}$ &  5.0 \\
\hline
\end{tabular}
\caption{Physical parameters and He$^+$/H$^+$ abundance solutions of the Final Dataset}
\label{table:GTO}
\end{table}
\end{landscape}

\section{Results from the Final Dataset} \label{Results}

Following the results of the previous sections, we now calculate the primordial helium abundance (mass fraction), $Y_{p}$, from the data listed in table~\ref{table:GTO}.  A regression of Y, the helium mass fraction, versus O/H, the oxygen to hydrogen mass fraction, is used to extrapolate to the primordial value\footnote{This work takes $Z=20(O/H)$ such that $Y=\frac{4y(1-20(O/H))}{1+4y}$}.  The O/H values are taken directly from ITS07\citep{its07}.  

\begin{figure}
\resizebox{\textwidth}{!}{\includegraphics{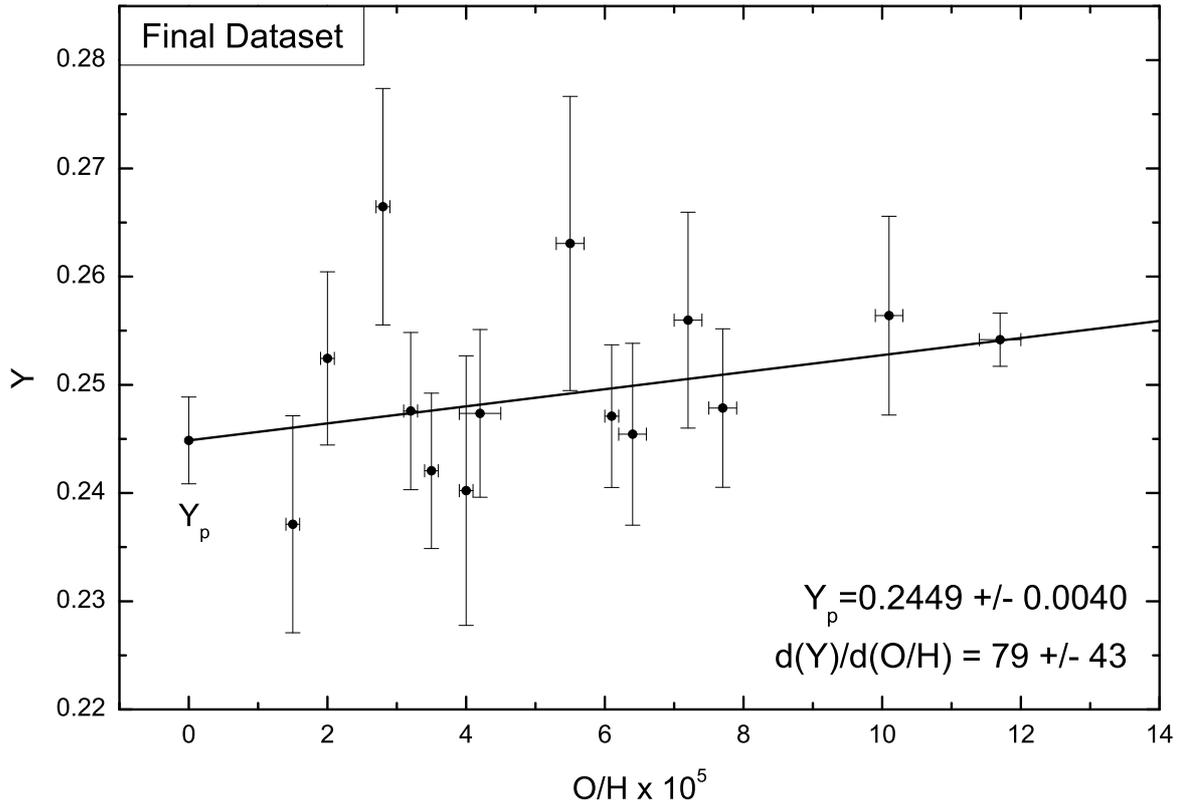}}
\caption{
Helium abundance (mass fraction) versus oxygen to hydrogen ratio regression calculating the primordial helium abundance.
}
\label{Y_OH}
\end{figure}

The relevant values for the Final Dataset's regression are given in table \ref{table:PH}.  The regression based on the 15 qualifying objects yields,
\beq
Y_p = 0.2449 \pm 0.0040,
\label{eq:Yp}
\eeq
with a slope of 79 $\pm$ 43 and a total  $\chi^2$ of 7.6.  The result is shown in figure \ref{Y_OH} and summarized in table \ref{table:Yp's}. This result for $Y_{p}$ agrees well with the SBBN value of $Y_p = 0.2471 \pm 0.0002$ \cite{cfoy}, based on the Planck determined baryon density \cite{planck15}.  Not surprisingly, eq.\ (\ref{eq:Yp}) also agrees well with the SBBN-independent, direct Planck estimation of $Y_p = 0.251 \pm 0.027$.  AOPS determined $Y_p = 0.2465 \pm 0.0097$ with a slope of 96 $\pm$ 122.  The result from AOPS is in good agreement with the newer result.  Restricting the qualifying AOPS dataset to the 11 objects which are shared with the qualifying dataset of this work, results in $Y_p = 0.2461 \pm 0.0107$ with slope 84 $\pm$ 144, again in good agreement with this result found in eq.\ (\ref{eq:Yp}).  Due to five more objects qualifying, the Final Dataset of this work contains a similar number of objects as in AOPS (16 there, 15 here).  However, as a result of the reduction in the individual $y^+$ uncertainties (see \S \ref{Sample}), the uncertainty on the intercept and slope are each reduced by approximately 60\%.  The intercept value decreased by 0.65\% compared to AOPS.  Given that the uncertainty on the intercept determination in AOPS was 3.9\%, this decrease is not significant.  

Including the 3 flagged objects\footnote{One of the flagged objects is another independent observation of SBS~0335-052E that is combined with its two qualifying observations.} decreases the intercept and reduces the uncertainty to $0.2424 \pm 0.0034$ with a slope of 116 $\pm$ 32.  The reduced uncertainty is primarily a result of the increased number of points in the regression, while the decreased intercept is \textit{entirely} the result of the increased slope.  As was seen in AOS3 \& AOPS, the flagged data points tend to have higher helium abundances, primarily due to the objects flagged for large neutral hydrogen fractions.  In previous analyses, where the evidence for a non-zero slope was marginal, we have reported the mean value of Y. The mean value of Y for the Final Dataset using only qualifying points is  $<$Y$>=0.2515 \pm 0.0017$. The mean value {\em increases} to $<$Y$>=0.2533 \pm 0.0016$, when the flagged objects are included.  \citet{os04} restricted the metallicity baseline to O/H $\le 9.2 \times 10^{-5}$.  Adopting the same metallicity cut with the dataset of this work increases the intercept to $0.2466 \pm 0.0063$, in near perfect agreement with the SBBN result using the Planck baryon density.  Note, however, that in this case the reduced baseline leaves us with an undetermined slope, and for this case alone, basing the primordial abundance on the mean value given in the table \ref{table:Yp's} is justified. 

Instead, if a regression analysis is performed excluding the T(O~III) prior, the 12 qualifying objects yield $Y_p = 0.2379 \pm 0.0050$.  However, the loss of the information provided by the T(O~III) prior and the reduced sample size weakens the value of this determination.  Nevertheless, the substantial reduction in the value of the intercept raises some questions regarding possible systematic errors when the prior is dropped.  

Starting with the 75 observations corresponding to the 45 H~II regions for which they observed near-infrared spectra to measure \10830, ITG14 select 28 for which $W(H\beta) \geq 150$ \AA, $O^{++}/O \geq 0.8$, and $\sigma(Y)/Y \leq 3\%$.  Their resulting regression analysis and accounting of the dominant systematic uncertainties finds $Y_p = 0.2551 \pm 0.0022$.  The significant differences between their result and the result derived here are due both to differences in determining the helium abundances for individual objects and in how the samples are defined.

\begin{table}[ht!]
\small
\centering
\vskip .1in
\begin{threeparttable}[t]
\begin{tabular}{lcccc}
\hline\hline
Object & 	He$^+$/H$^+$ 	      & He$^{++}$/H$^+$     & Y 		  & O/H $\times$ 10$^5$ \\
\hline
\multicolumn{5}{c}{Final Dataset Not Flagged (Qualifying)} \\
\hline
I~Zw~18~SE~1	&	0.07693	$\pm$	0.00423	&	0.0008	$\pm$	0.0008	&	0.2371	$\pm$	0.0100	&	1.5	$\pm$	0.1	\\
SBS~0335-052E1+3\tnote{(a)}	&	0.08201	$\pm$	0.00303	&	0.0026	$\pm$	0.0018	&	0.2524	$\pm$	0.0080	&	2	$\pm$	0.1	\\
J0519+0007	&	0.08875	$\pm$	0.00461	&	0.0021	$\pm$	0.0021	&	0.2665	$\pm$	0.0109	&	2.8	$\pm$	0.1	\\
SBS~0940+544~2	&	0.08179	$\pm$	0.00316	&	0.0005	$\pm$	0.0005	&	0.2476	$\pm$	0.0073	&	3.2	$\pm$	0.1	\\
Tol~65	&	0.07883	$\pm$	0.00294	&	0.0011	$\pm$	0.0011	&	0.2421	$\pm$	0.0072	&	3.5	$\pm$	0.1	\\
SBS~1415+437~(No.~1)~3	&	0.07694	$\pm$	0.00494	&	0.0022	$\pm$	0.0022	&	0.2402	$\pm$	0.0125	&	4	$\pm$	0.1	\\
SBS~1415+437~(No.~2)	&	0.08226	$\pm$	0.00343	&	0.0000	$\pm$	0.0000	&	0.2474	$\pm$	0.0078	&	4.2	$\pm$	0.3	\\
CGCG~007-025~(No.~2)	&	0.08867	$\pm$	0.00623	&	0.0007	$\pm$	0.0007	&	0.2631	$\pm$	0.0136	&	5.5	$\pm$	0.2	\\
Mrk~209	&	0.08114	$\pm$	0.00272	&	0.0011	$\pm$	0.0011	&	0.2471	$\pm$	0.0066	&	6.1	$\pm$	0.1	\\
SBS~1030+583	&	0.07937	$\pm$	0.00306	&	0.0021	$\pm$	0.0021	&	0.2454	$\pm$	0.0084	&	6.4	$\pm$	0.2	\\
Mrk~71~(No.~1)	&	0.08539	$\pm$	0.00445	&	0.0008	$\pm$	0.0008	&	0.2560	$\pm$	0.0100	&	7.2	$\pm$	0.2	\\
SBS~1152+579	&	0.08139	$\pm$	0.00303	&	0.0012	$\pm$	0.0012	&	0.2478	$\pm$	0.0073	&	7.7	$\pm$	0.2	\\
Mrk~59	&	0.08548	$\pm$	0.00405	&	0.0010	$\pm$	0.0010	&	0.2564	$\pm$	0.0092	&	10.1	$\pm$	0.2	\\
SBS~1135+581	&	0.08462	$\pm$	0.00072	&	0.0008	$\pm$	0.0008	&	0.2542	$\pm$	0.0025	&	11.7	$\pm$	0.3	\\
Mrk~450~(No.~1)	&	0.08634	$\pm$	0.00441	&	0.0003	$\pm$	0.0003	&	0.2565	$\pm$	0.0097	&	15.2	$\pm$	0.4	\\
\hline
\multicolumn{5}{c}{Final Dataset with Flags} \\
\hline
SBS~0335-052E1+2+3\tnote{(a)}	&	0.08132	$\pm$	0.00243	&	0.0025	$\pm$	0.0015	&	0.2509	$\pm$	0.0064	&	2	$\pm$	0.1	\\
Mrk~1315	&	0.09341	$\pm$	0.00274	&	0.0000	$\pm$	0.0000	&	0.2710	$\pm$	0.0058	&	18.9	$\pm$	0.4	\\
Mrk~1329	&	0.09060	$\pm$	0.00425	&	0.0000	$\pm$	0.0000	&	0.2650	$\pm$	0.0091	&	19.2	$\pm$	0.5	\\
\hline
\end{tabular}
\begin{tablenotes}
  \item [(a)] SBS~0335-052E1, SBS~0335-052E2, \& SBS~0335-052E3 are independent observations of the same H~II region.  As such, they were combined by weighted average into a single regression point.
\end{tablenotes}
\end{threeparttable}
\caption{Primordial helium regression values}
\label{table:PH}
\end{table}

\begin{table}[ht!]
\small
\centering
\vskip .1in
\begin{threeparttable}[t]
\begin{tabular}{lcccc}
\hline\hline
Dataset							& N    	 & Y$_p$	 	& dY/d(O/H)	& $<$Y$>$ \\
\hline
Qualifying						& 15	& 0.2449 $\pm$ 0.0040	&  79 $\pm$  43	& 0.2515 $\pm$ 0.0017 \\
Qualifying + Flagged					& 17	& 0.2424 $\pm$ 0.0034	& 116 $\pm$  32	& 0.2533 $\pm$ 0.0016 \\
Qualifying\tnote{(a)}~~\,, $O/H < 9.2 \times 10^{-5}$		& 12	& 0.2466 $\pm$ 0.0063	&  35 $\pm$ 125	& 0.2482 $\pm$ 0.0025 \\
Qualifying, T(O~III) prior removed			& 12	& 0.2379 $\pm$ 0.0050	& 145 $\pm$  52	& 0.2508 $\pm$ 0.0019 \\
\hline
\end{tabular}
\begin{tablenotes}
  \item [(a)] \citet{os04} restricted the metallicity baseline to O/H $\le 9.2 \times 10^{-5}$, and the same dataset was analyzed in AOS \& AOS2 \citep{AOS,AOS2}.
\end{tablenotes}
\end{threeparttable}
\caption{Comparison of Y$_p$ for selected datasets}
\label{table:Yp's}
\end{table}

\section{Discussion} \label{Conclusion}

The primary aim of this work was to evaluate the effects of adding \10830 to a helium abundance analysis based solely on optical emission lines.  In particular, it was hoped that this emission line, due to its much stronger electron density dependence, would have a significant effect in constraining the density in our solutions and, thereby, effectively break the degeneracy between temperature and density.  Preliminary synthetic testing and analysis of H~II region observations showed  the positive impact of \10830 in improving the determination of our solution parameters, $n_e$, $T_e$, and of primary interest, $y^+$.  Most objects saw their $y^+$ uncertainty decrease by $\sim$50\%, with a corresponding, similar improvement in the precision of $Y_p$.  

Our determination of $Y_p = 0.2449 \pm 0.0040$ is in good agreement with the SBBN predicted value (as well as our previous results), and while the value of the intercept has decreased, it differs from the SBBN prediction by approximately 1/2 $\sigma$, even with its most welcome increased precision. There are, however, potential systematic effects to be addressed. First, the inclusion of the near-infrared \10830 with existing optical spectra requires scaling of \10830 (see \S \ref{ITG14}).  This naturally raises the possibility of introducing additional error or systematic bias.  Second, as discussed in \citet{pfms} and updated in \citet{pfsd}, the helium emissivities, the primary driver in determining the helium abundance from the observed fluxes, carry estimated uncertainties of $\sim$0.2-0.7\%.  Furthermore, significant, systematic improvements in the helium emissivities have been made in recent years \citep{pfm07, pfsd}.  Any further shifts in these emissivities will have a significant effect on the determination of $Y_p$.  

The impressive benefits from including \10830 in nebular helium abundance determinations strongly supports the case for additional near-infrared spectral observations of low-metallicity H~II regions.  In particular, to remove potential systematic effects introduced by scaling \10830 observations, simultaneous observations of the optical and infrared spectra would be decisive.  Coupled with the opportunity for higher quality, higher resolution spectra, collection of simultaneous optical and infrared spectra would be of considerable value and utility.  In sum, there is potential to significantly reduce the uncertainty in the determination of $Y_p$. 

\acknowledgments

We would like to thank Sergio S\'imon-D\'iaz, Max Pettini, and Ryan Cooke for valuable conversations. This work has made use of NASA's Astrophysics Data System Bibliographic Services and the NASA/IPAC Extragalactic Database (NED), which is operated by the Jet Propulsion Laboratory, California Institute of Technology, under contract with the National Aeronautics and Space Administration.

The work of KAO is supported in part by DOE grant DE-SC0011842.  EDS is grateful for partial support from the University of Minnesota.


\end{document}